\documentclass[lettersize,journal]{IEEEtran}
\usepackage{amsmath,amsfonts}
\usepackage{algorithmic}
\usepackage{algorithm}
\usepackage{array}
\usepackage{textcomp}
\usepackage{stfloats}
\usepackage{url}
\usepackage{verbatim}
\usepackage{graphicx}
\usepackage{cite}
\usepackage{hyperref}
\usepackage{subfigure}
\graphicspath{{./Figures/}}

\begin{document}

\title{\fontsize{24}{28}\selectfont Mechanically Reconfigurable GRIN Lens Concept for Focusing and Beamforming Applications}


\author{
	\IEEEauthorblockN{K. Kaboutari\IEEEauthorrefmark{1}\IEEEauthorrefmark{2},X. Liu\IEEEauthorrefmark{2}, A. Abraray\IEEEauthorrefmark{1}, P. Pinho\IEEEauthorrefmark{1}, S. Shen\IEEEauthorrefmark{2}, S. Maslovski\IEEEauthorrefmark{1}}
	
\IEEEauthorblockA{\IEEEauthorrefmark{1} Instituto de Telecomunica\c{c}\~{o}es and Dept. de Eletrónica, Telecomunica\c{c}\~{o}es e Inform\'{a}tica, University of Aveiro, Aveiro, Portugal\\
	\{k.kaboutari, abdelghafour.abraray, ptpinho, stanislav.maslovski\}\IEEEauthorrefmark{1}@ua.pt}

\IEEEauthorblockA{\IEEEauthorrefmark{2}Department of Mechanical Engineering, Carnegie Mellon University, Pittsburgh, PA 15213, United States of America\\
	\{kkabouta, xiuliu, shengshe\}\IEEEauthorrefmark{2}@andrew.cmu.edu}
	}



\pagenumbering{gobble}

\maketitle

\begin{abstract}
    This study presents a prospective concept of a mechanically reconfigurable Gradient Index (GRIN) lens for focusing and beamforming applications. The lens is formed by corrugated layers, each realizing a refractive index profile expressed in a basis of Chebyshev polynomials. The focal point position is controlled by shifting the GRIN lens layers. A geometrical optics-based approach is implemented for the three scenarios of focusing at varying focal distance. The geometrical optics results are validated by numerical simulations, with some discrepancies observed due to modeling approximations and structural granularity. We also discuss scalability of the proposed concept for operation at infrared wavelengths.
\end{abstract}

\begin{IEEEkeywords}
    Reconfigurable GRIN Lens, Focusing, Beamforming, Optimization.
\end{IEEEkeywords}

\section{Introduction}
    \IEEEPARstart{L}{enses} are essential for focusing incident Electromagnetic (EM) waves and forming images of objects. They have been studied at microwave frequencies~\cite{Kock_1946, Goodman_2005, Peeler_1953} for various applications in radar systems~\cite{Raman_1998, Tang_2021}, telecommunications~\cite{Chen_2011, Meng_2018}, non-destructive testing~\cite{Gopalsami_1994}, and imaging~\cite{Datta_2022} as well as in the optical range of the EM spectrum~\cite{Hanninen_2019}. Historically, microwave lenses followed the optical approach of shaping dielectric materials into convex or concave bodies of revolution. However, such conventional approaches are associated with bulky and cumbersome profiles and costly machining processes.
    
    Gradient Index (GRIN) lenses are planar optical lenses with spatially varying refractive index / permittivity. They offer a fix to conventional lens problems such as bulkiness, complicated profiles, and high fabrication costs. They can be realized as engineered structures composed of a periodic arrangement of subwavelength elements, known as unit cells or meta-atoms, which are significantly smaller than the operating wavelength $\lambda$. According to the effective medium theory, such an arrangement can be described by effective permittivity and permeability. The unit cell size must be less than $\lambda/4$ to ensure these inhomogeneous periodic materials behave like a homogeneous medium for incident waves~\cite{Slovick_2014, Makdissy_2016, Alu_2011}.
    
    The bulk EM properties of these lenses are determined by the configuration and content of the unit cells rather than the atomic-scale composition typical of dielectric materials. In fact, GRIN lenses use a refractive index / permittivity gradient within the material to bend and focus EM waves, unlike conventional lenses that rely on curved surfaces to focus EM waves in accordance with Snell's law. This gradual bending of rays occurs because the waves propagate at different phase velocities in regions of different permittivities. The varying permittivity induces a parabolic trajectory for the rays, allowing them to converge or diverge smoothly, thus facilitating focusing or collimation of the EM waves. This feature offers several advantages, particularly in applications where precision and compactness of the lenses are critical for efficient operation. Consequently, GRIN lenses can be designed to exhibit specialized EM properties that enable novel lens applications that are challenging or impossible to achieve with conventional techniques. These features allow GRIN lenses to deliver comparable or even superior focusing capabilities within a more compact physical structure.
    
    In~\cite{Datta_2022}, a GRIN metasurface lens is designed for microwave imaging at 8~GHz. The metasurface's unit cell comprises an electric-LC (ELC) resonator, through which the metasurface's effective refractive index is manipulated by adjusting the capacitive gap at the unit cell's center, enabling the creation of a GRIN layer. Subsequently, a horn antenna serves as a source of plane waves incident on the lens for assessing the lens's focusing capabilities.
    In~\cite{Chaky_2023}, it is shown that placing a GRIN lens on top of a transmitting horn antenna effectively redistributes the field across the reflector's surface. This redistribution reduces the incident field strength variation across the reflecting aperture by 17.6\% as compared to the nominal case without the lens enclosure. This coincides with a substantial 47\% increase in the power handling of the system. The increase in power handling has practical implications, suggesting that the GRIN-enhanced system can either sustain a stronger incident field or use a more powerful source at the transmitter side, potentially enhancing the system's performance.
    
    A reconfigurable GRIN lens for biomedical applications proposed in~\cite{Rezaeieh_2011} offers a promising solution to the challenges faced in radar-based imaging algorithms. It enhances the accuracy of the imaging algorithms by converting spherical wavefronts into plane waves. This is a significant step forward, considering the limitations of the current body-matched antennas that do not produce plane waves. The lens is expected to reduce errors in the obtained images, particularly in imaging cancerous tissues to determine the location of the malignancy, its size, and prevent formation of ghost targets, which are often a result of the plane wave assumption of the algorithms. In their design, a reconfigurable GRIN lens is achieved by filling cavities within the lens with water, which changes the local effective permittivity of the lens. However, this approach requires using bulky equipment such as water tanks and pumps, which may be challenging or even impossible in certain cases.
    
    Another research presented in~\cite{Allen_2013} focuses on a metamaterial implementation using 3D printing of a GRIN lens with radially varying refractive index gradient and without polarization constraints. This implementation is specifically designed to enhance the directivity of beam scanning in X-band microwave antenna systems. The designed GRIN lens offers significant advantages, providing a new level of freedom in manipulating electromagnetic waves while maintaining optimal performance. The GRIN lens's permittivity is estimated using the effective medium theory. The optimized design was then used to fabricate the GRIN lens with an isotropic, inhomogeneous dielectric material, in which the refractive index is carefully designed to match the theoretical results using a mixture ratio of air-filled voids and a polymer. Although practically any refractive index profile can be realized through this approach, the produced 3D-printed lens is not reconfigurable as such. 
    
    In contrast to the studies discussed above, our study presents a novel working principle for mechanically reconfigurable laminated / layered GRIN lenses with controllable refractive index profiles for focusing and beamforming applications at microwave, millimeter-wave, and, possibly, near-infrared bands. The layers are formed by different materials, each with an appropriate refractive index. Corrugations or grooves with pre-designed patterns corresponding to appropriate basis functions are then etched in the layers. This approach allows the relative permittivity of the layers to change independently, producing the desired effective refractive index / relative permittivity profile when all basis functions are combined together. The overall effective refractive index profile is then controlled by lateral shifts of the layers, in contrast to axial shifts as in conventional optical systems. We foresee potential applications of this technique in the fields of biomedical imaging and telecommunications.
\section{Methodology}
\label{Methodology of the Proposed Approach}
    In our methodology, the GRIN lens is partitioned into multiple layers, which provides flexibility to assign different materials to each layer and independently shift the layers mechanically. Beamforming is then achieved by creating suitable refractive index profiles across the layers and adjusting their positions accordingly.
    
\begin{figure}[ht!]
        \begin{center}
            \subfigure[]{\label{Fig: Fig01.a}
            \includegraphics[width=1\linewidth,trim={0.0cm 0.0cm 0.0cm 0.0cm},clip=true]{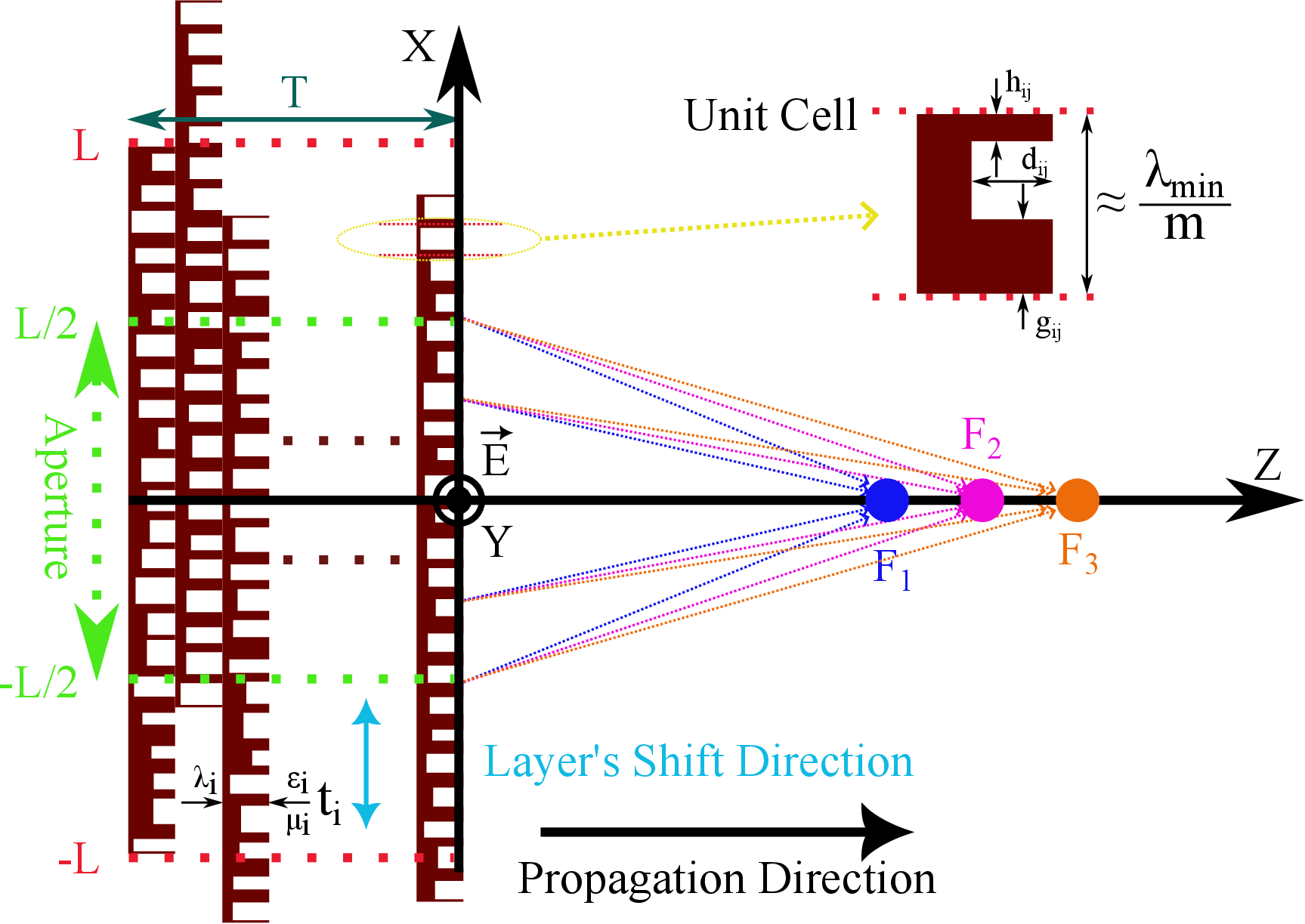}}
            
            \subfigure[]{\label{Fig: Fig01.b}
            \includegraphics[width=0.9\linewidth,trim={0.0cm 0.0cm 0.0cm 0.0cm},clip=true]{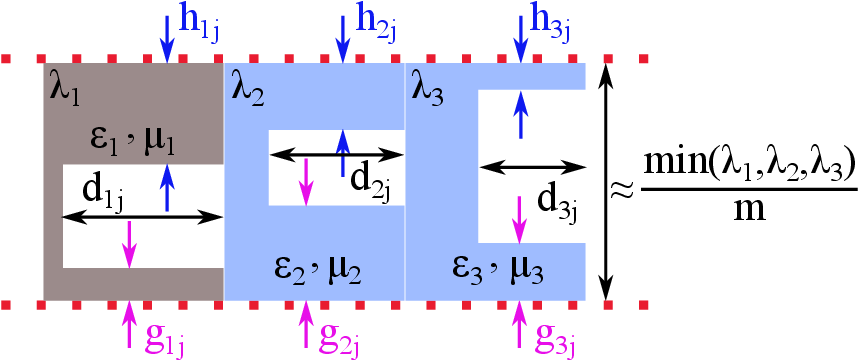}}
        \end{center}
        \vspace{-3.2mm}
        \caption{Geometry of the proposed GRIN lens configuration: (a) Structural schematics and the geometrical optics ray picture; (b) Supercell of the GRIN lens. $\varepsilon_{i}$ and $\mu_{i}$ are the relative permittivity and permeability of the materials. Here, $i$ and $j$ represent layer and groove indices, respectively.}
        \label{Fig: Fig01}
    \end{figure}

    In this study, we use patterns of air-filled grooves with different depths and heights inside the layers to control the transmitted field phase and concentrate the field at a set of focal points, Fig~\ref{Fig: Fig01.a}. The GRIN lens depicted in this figure consists of multiple magneto-dielectric layers arranged along the $z$-axis. Each layer features independent materials with different thickness and refractive index (i.e., relative permittivity and permeability).
    The grooves are aligned with the $y$-axis, which is also the polarization direction of the incident electric field, thereby avoiding induction of dielectric polarization charges on the corrugate boundaries and the host medium. Moreover, in this study, in order to simplify the comparison of analytical and numerical results, the characteristic impedance of the materials, $\eta_{i}$, was adjusted such that $\eta_{i}$ equals $\eta_{0}=\sqrt{\frac{\mu_{i, \rm eff}\mu_{0}}{\varepsilon_{i, \rm eff}\varepsilon_{0}}}\approx 377\ \Omega$. Such impedance matching to free space can be achieved by using magneto-dielectric materials, or by adding impedance-matching layers (e.g., impedance grids or metasurfaces) between the dielectric layers. 
    
    The illustration in Fig~\ref{Fig: Fig01.a} also depicts a unit cell for one of the layers. The height of the unit cell in the $x$-axis should be at most 1/4 of the operating wavelength. The $m$ value is selected to be 10 in our research. The specific values for $h_{ij}$, $d_{ij}$, and $g_{ij}$ are established based on the necessary relative permittivity/refractive index of the layer at the unit cell's position.
    
    In the optimization studies, we chose to design a GRIN lens using three different magneto-dielectric substrates. A supercell of the proposed configuration is depicted in Fig.~\ref{Fig: Fig01.b}. In this structure, the first layer has a refractive index of $\sqrt{10.2}$. The second and third layers have the refractive index of $\sqrt{6.15}$ (these values were selected to match the refractive indices of Rogers RO3010 and RO3006 substrates, respectively).
    
    In this design, the grooves in the first mechanically fixed layer are created so that its effective refractive index profile follows the Chebyshev polynomial of the second-order. On the other hand, the second and third layers of our GRIN lens can be mechanically shifted along the $x$-axis in opposite directions with respect to each other. The groove profiles for the second and third layers are determined using Chebyshev polynomials of third-order mutually mirrored with respect to $x=0$ to present maximum and minimum values for relative permittivity. Therefore, the lens's effective relative permittivity can be determined using the superposition (i.e. mixing rule) concept for the relative permittivity of the layers.
    
    As was mentioned above, the thickness of each GRIN lens layer should be small (typically less than $\lambda_{i}/4$) in order to exhibit characteristics of a homogenized medium when interacting with incident waves. However, we found that when the layers are impedance matched, these conditions can be relaxed without deteriorating the GRIN lens performance too much, as will be discussed in the next section.
    
\section{Basis Functions and Layer Patterning for Tailored Refractive Index Profiles and Multi-Focal Caustic}
\label{Basis Functions}
\begin{figure*}[ht!]
        \begin{center}
            \subfigure[]{\label{Fig: Fig02.a}
            \includegraphics[width=0.45\linewidth,trim={0.05cm 0.0cm 0.1cm 0.05cm},clip=true]{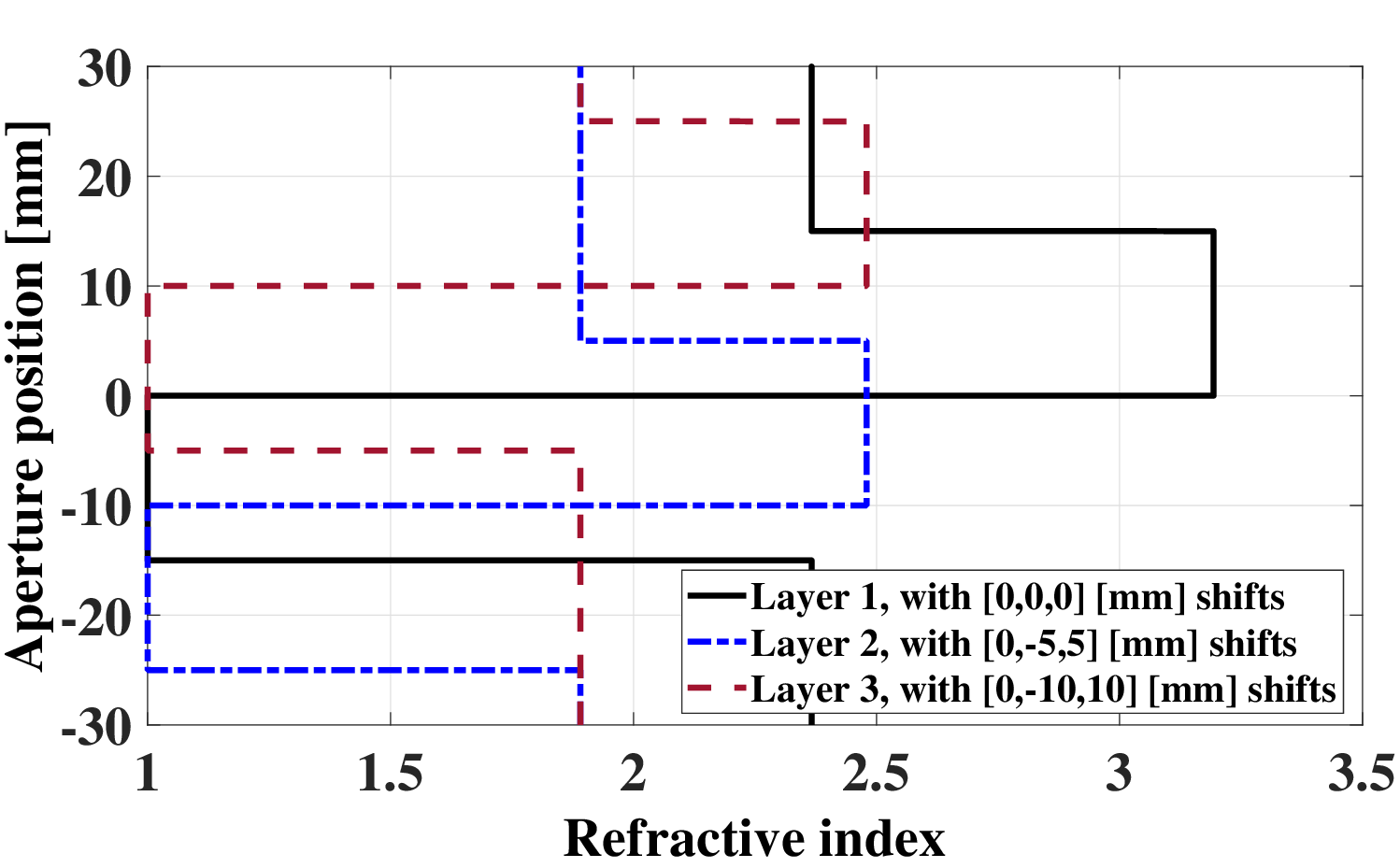}}
            \subfigure[]{\label{Fig: Fig02.b}
            \includegraphics[width=0.45\linewidth,trim={0.05cm 0.0cm 0.1cm 0.05cm},clip=true]{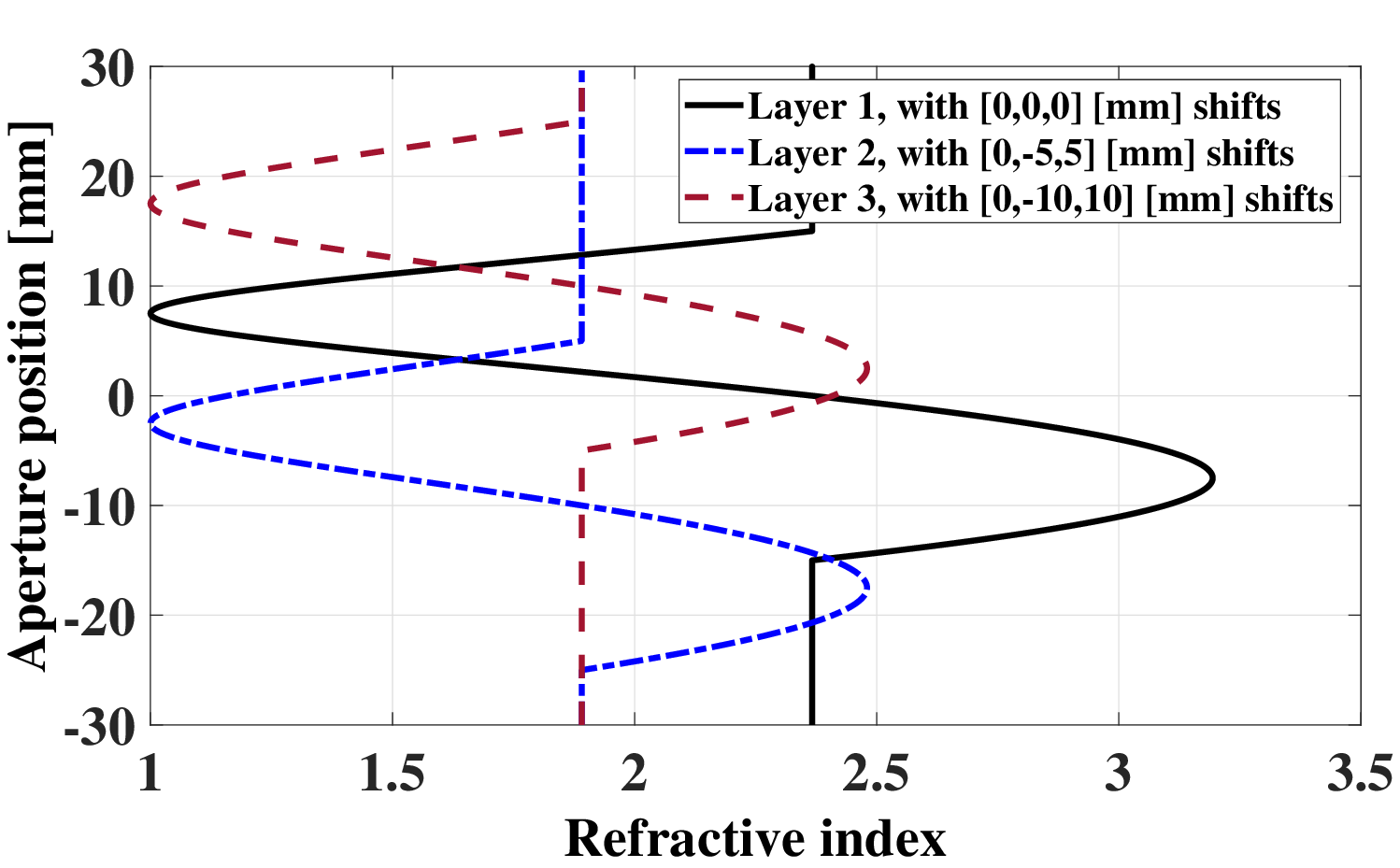}}
            \vspace{-3.5mm}
             \subfigure[]{\label{Fig: Fig02.c}
            \includegraphics[width=0.45\linewidth,trim={0.05cm 0.0cm 0.1cm 0.05cm},clip=true]{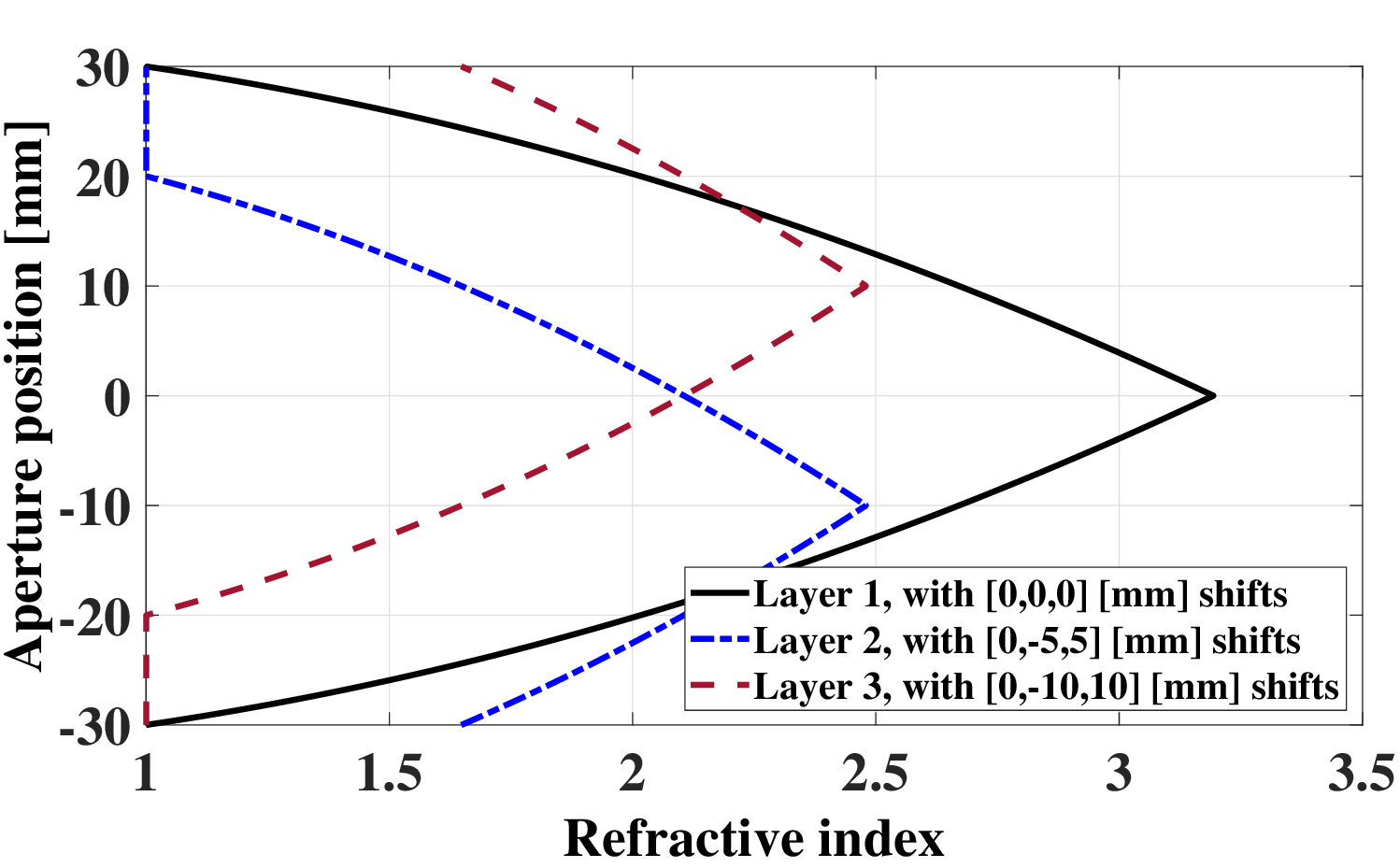}}
             \subfigure[]{\label{Fig: Fig02.d}
            \includegraphics[width=0.45\linewidth,trim={0.05cm 0.0cm 0.1cm 0.05cm},clip=true]{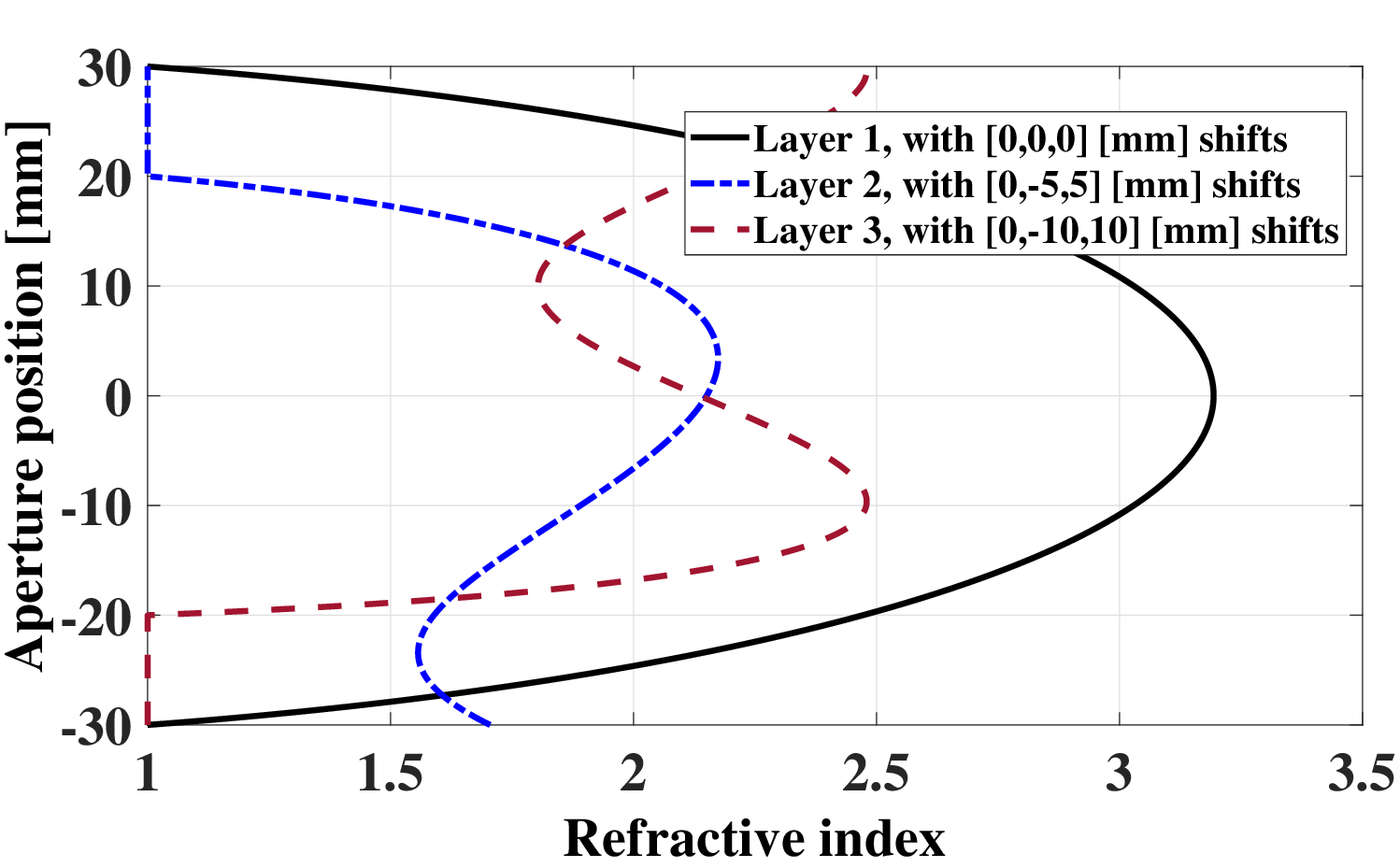}}
            \vspace{-3.5mm}
             \subfigure[]{\label{Fig: Fig02.e}
            \includegraphics[width=0.45\linewidth,trim={0.05cm 0.0cm 0.1cm 0.05cm},clip=true]{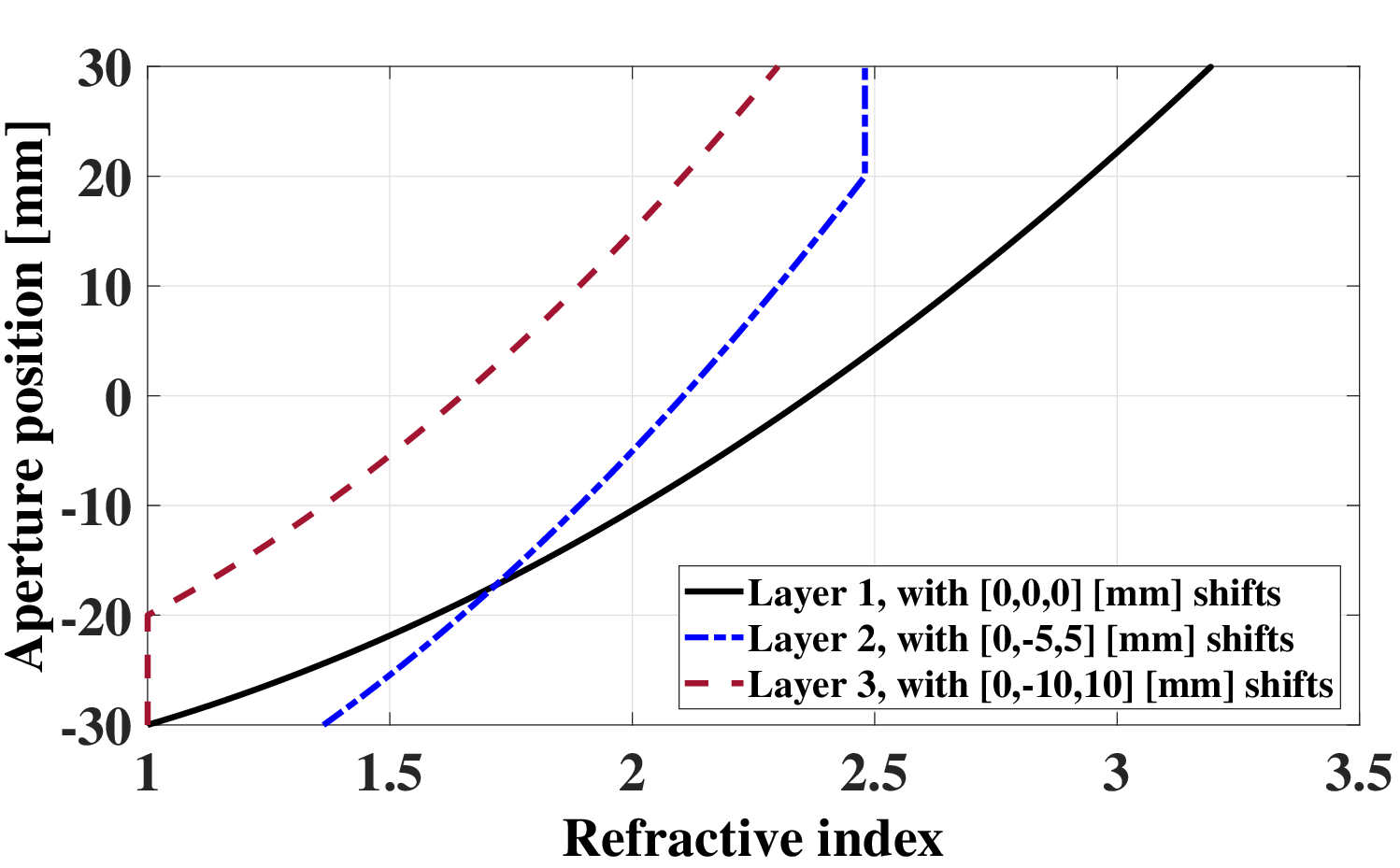}}
             \subfigure[]{\label{Fig: Fig02.f}
            \includegraphics[width=0.45\linewidth,trim={0.05cm 0.0cm 0.1cm 0.05cm},clip=true]{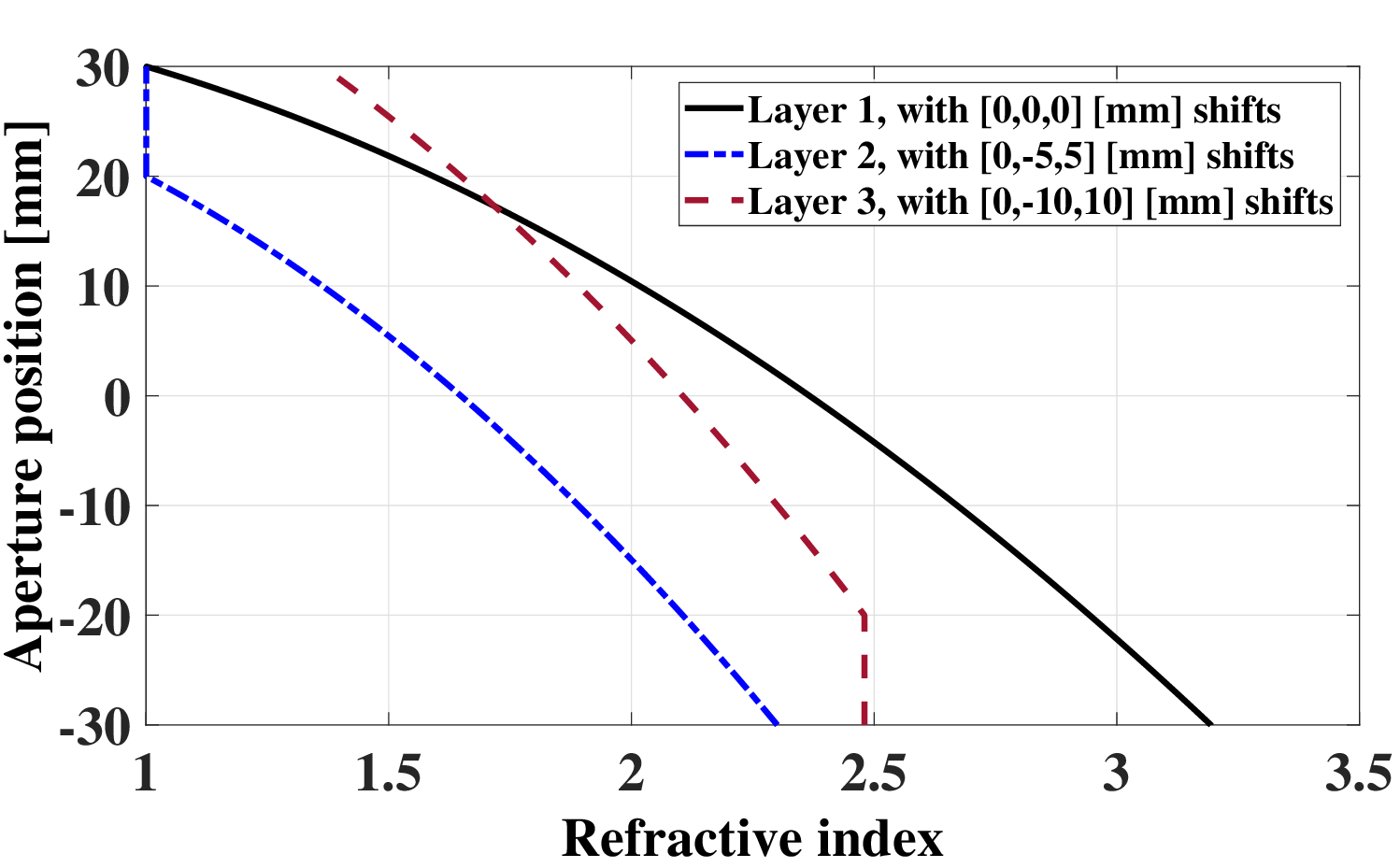}}
            \vspace{-3.5mm}
             \subfigure[]{\label{Fig: Fig02.g}
            \includegraphics[width=0.45\linewidth,trim={0.05cm 0.0cm 0.1cm 0.05cm},clip=true]{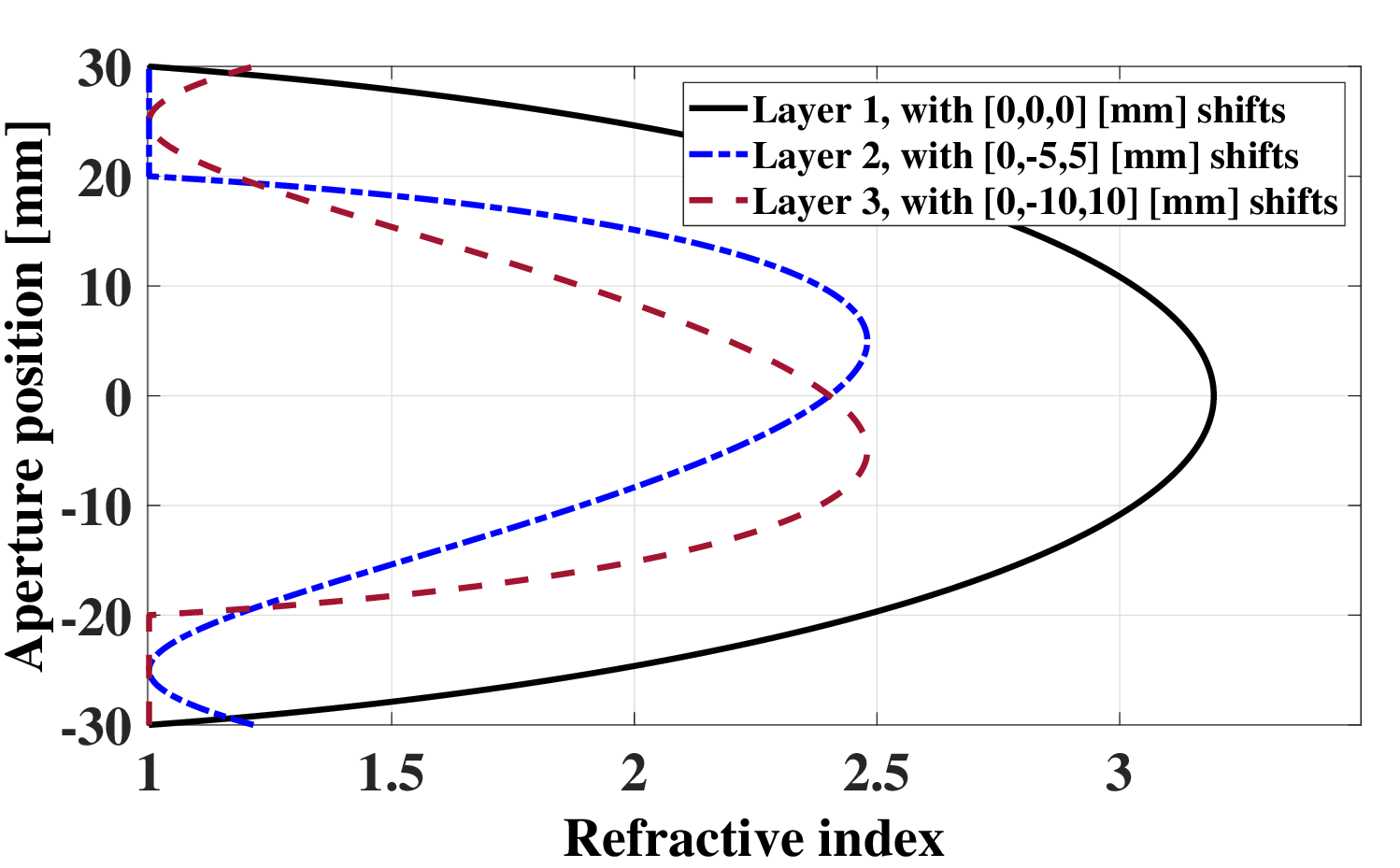}}
        \end{center}
        \vspace{-3.5mm}
        \caption{Refractive index profiles of the GRIN lens layers with the relative permittivity distributed as: (a) Haar wavelets, (b) Sinusoidal, (c) Sawtooth, (d) Legendre polynomials, (e) Linearly increasing, (f) Linearly decreasing, and (g) first-kind Chebyshev polynomials for the three cases of mutual shifts.}
        \label{Fig: Fig02}
    \end{figure*}

    Selecting appropriate basis functions and etching grooves with varying depths in each layer are essential for achieving the desired effective refractive index profile along the $x$-axis, Fig.~\ref{Fig: Fig03.a}. As a result of this patterning, different focal points can be created along the $z$-axis (i.e., the main direction of propagation) by shifting the layers independently and deliberately along the $x$-axis. The choice of basis functions and layer shifts is critical and should be assessed using optimization methods, which are instrumental in ensuring the creation of the intended refractive index profile along the $x$-axis. In our study, we examined various polynomial basis functions and their ability to approximate and generate smooth variations in the refractive index of the layers using optimization algorithms. These included Haar wavelets, sinusoidal, sawtooth, Legendre polynomials, linear, and first-kind Chebyshev polynomials basis functions. According to Fig.~\ref{Fig: Fig02}, the square root of the relative permittivity profile (refractive index profile) of the discussed basis functions can be observed. The selection of first-kind Chebyshev polynomials as the primary functional basis was a significant discovery, providing a clear direction for further research and application.
    
\section{Analytical and Numerical Results}
\label{Analytical and Numerical Results}
\begin{figure}[ht!]
        \begin{center}
            \subfigure[]{\label{Fig: Fig03.a}
            \includegraphics[width=1\linewidth,trim={0.05cm 0.05cm 0.15cm 0.05cm},clip=true]{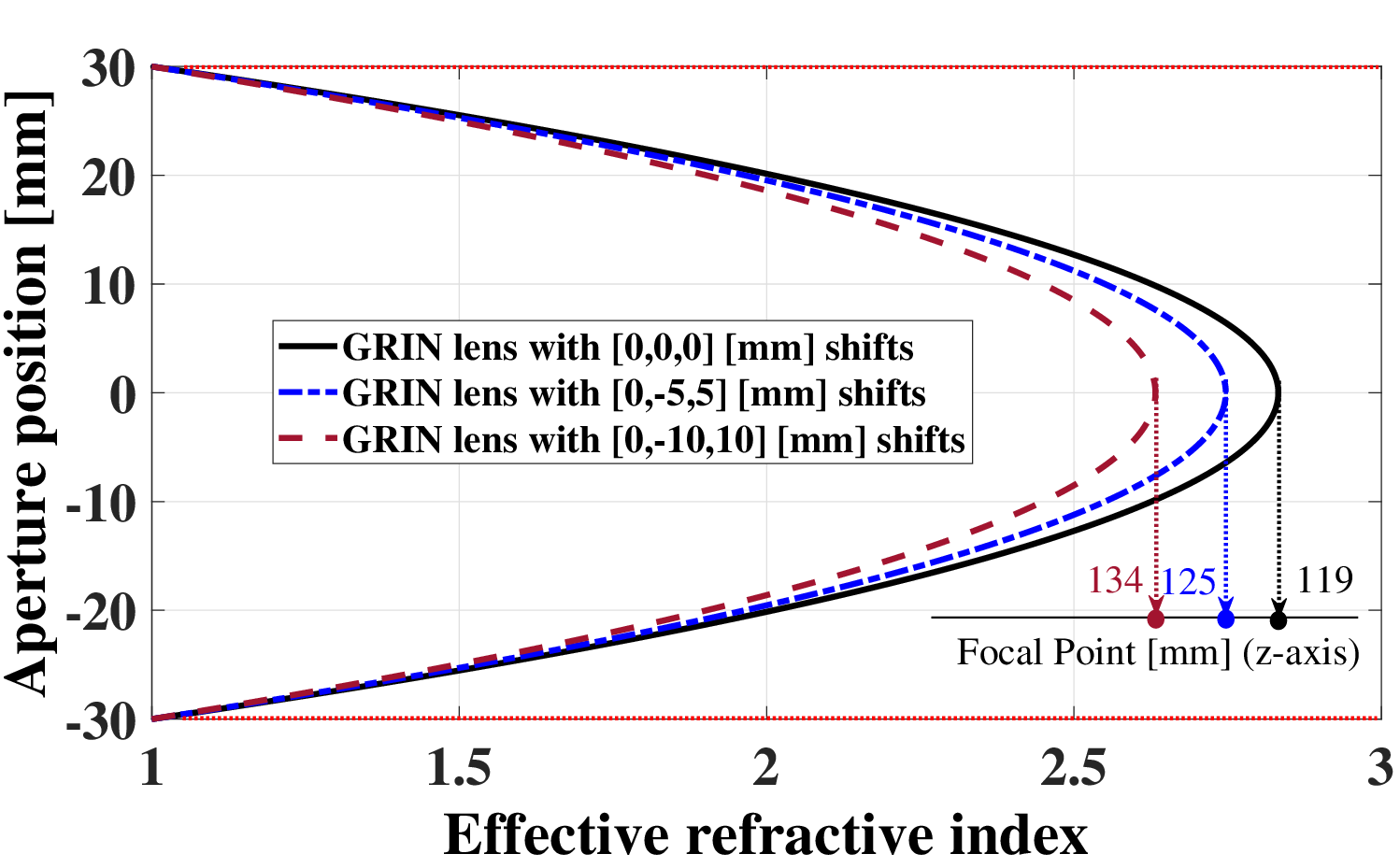}}
            
            \subfigure[]{\label{Fig: Fig03.b}
            \includegraphics[width=1\linewidth,trim={0.05cm 0.05cm 0.15cm 0.05cm},clip=true]{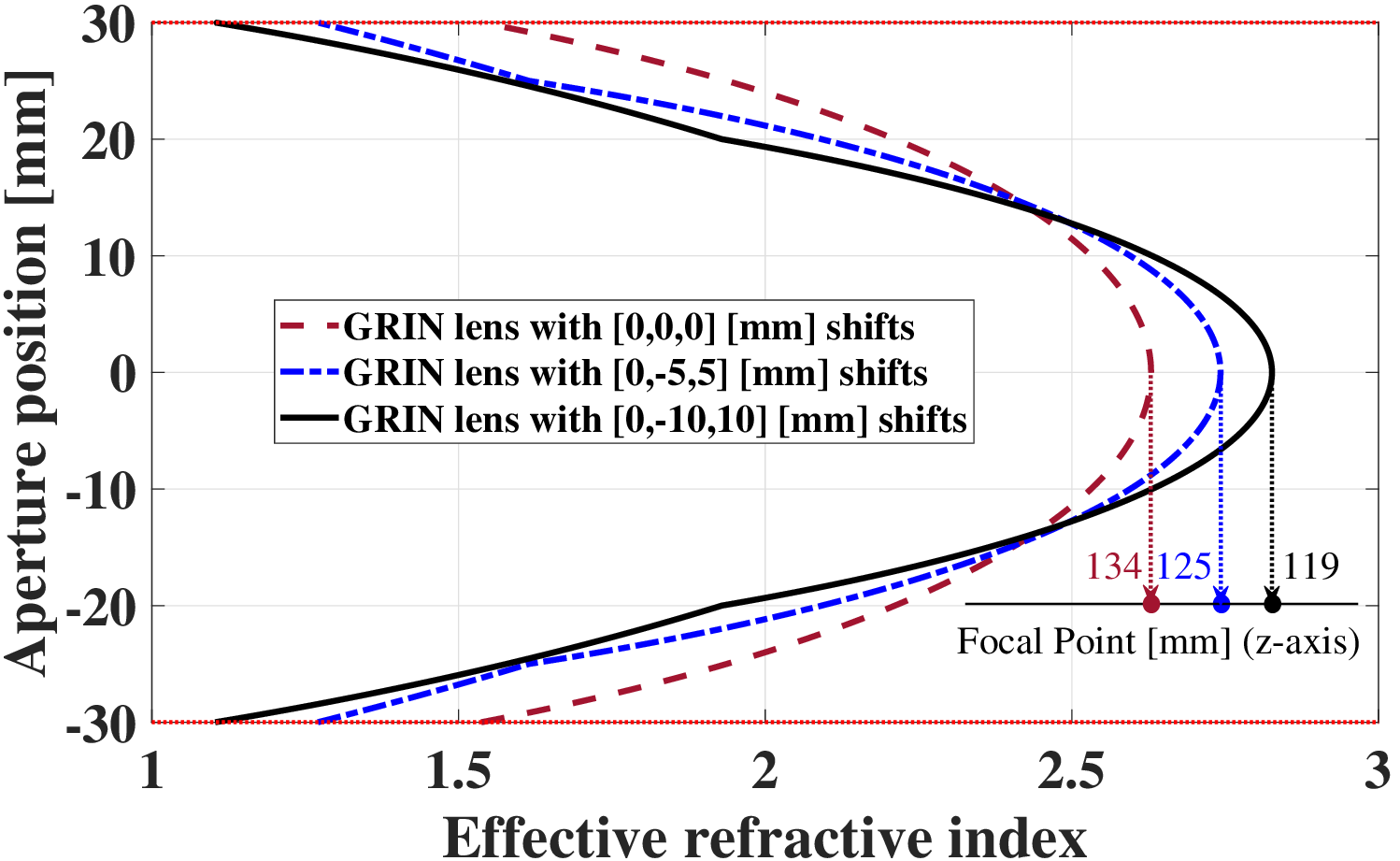}}
        \end{center}
        \vspace{-3.2mm}
        \caption{Effective refractive index profile and focal points: (a) An ideal profile for the three cases of mutual shifts, and (b) A 3-layer GRIN lens with layers of 1.02, 0.508, and 0.508~mm thickness for the three cases of mutual shifts.}
        \label{Fig: Fig03}
    \end{figure}

    In our analytical geometric optics-based model we have considered three scenarios to focus rays at 134~mm, 125~mm, and 119~mm along the $z$-axis using the proposed 3-layer GRIN lens. To achieve this, we have applied $[0,0,0]$, $[0,-5,+5]$, and $[0,-10,+10]$~mm shifts on the corresponding layers to create the intended effective refractive index profiles for the proposed reconfigurable GRIN lens, as shown in Fig.~\ref{Fig: Fig03.b}.
    
\begin{figure}[ht!]
        \begin{center}
            \subfigure[]{\label{Fig: Fig04.a}
            \includegraphics[width=1\linewidth,trim={0.45cm 0.0cm 1.4cm 0.5cm},clip=true]{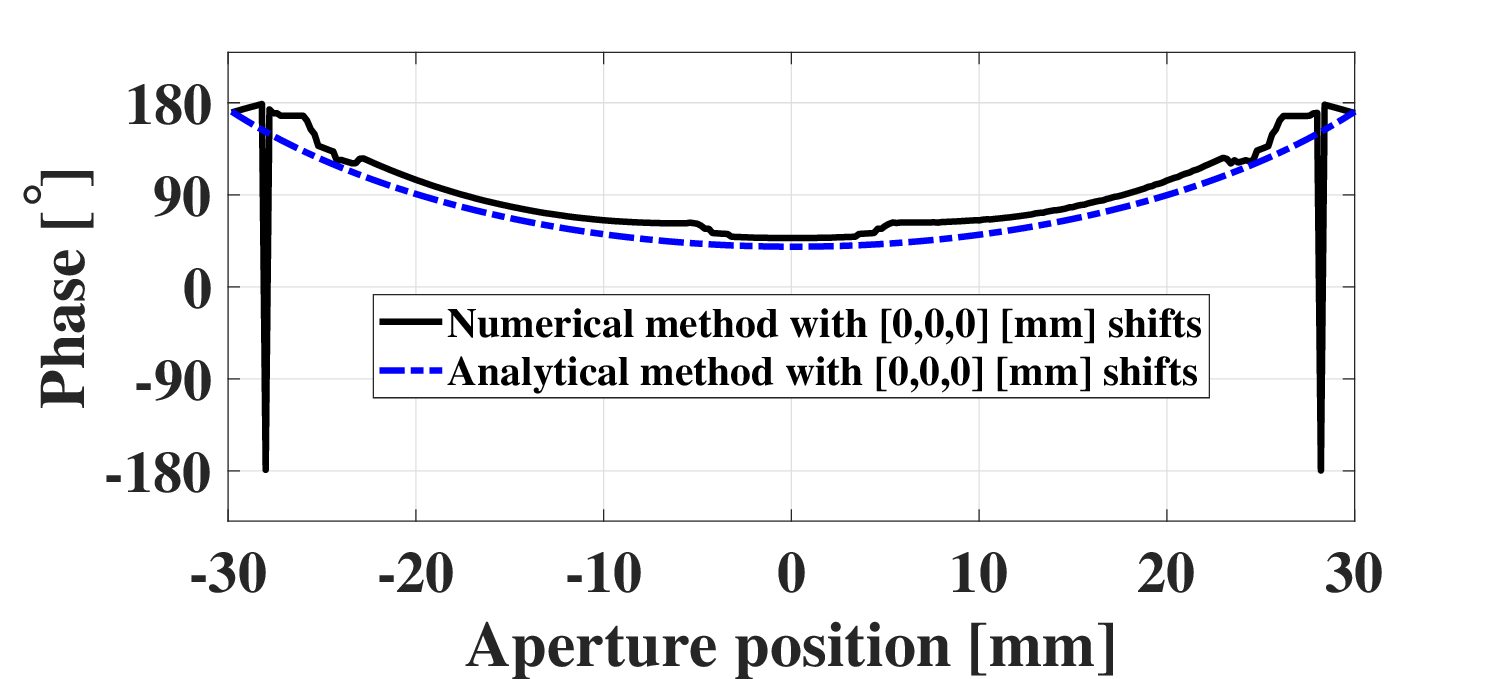}}
            
            \subfigure[]{\label{Fig: Fig04.b}
            \includegraphics[width=1\linewidth,trim={0.45cm 0.0cm 1.4cm 0.5cm},clip=true]{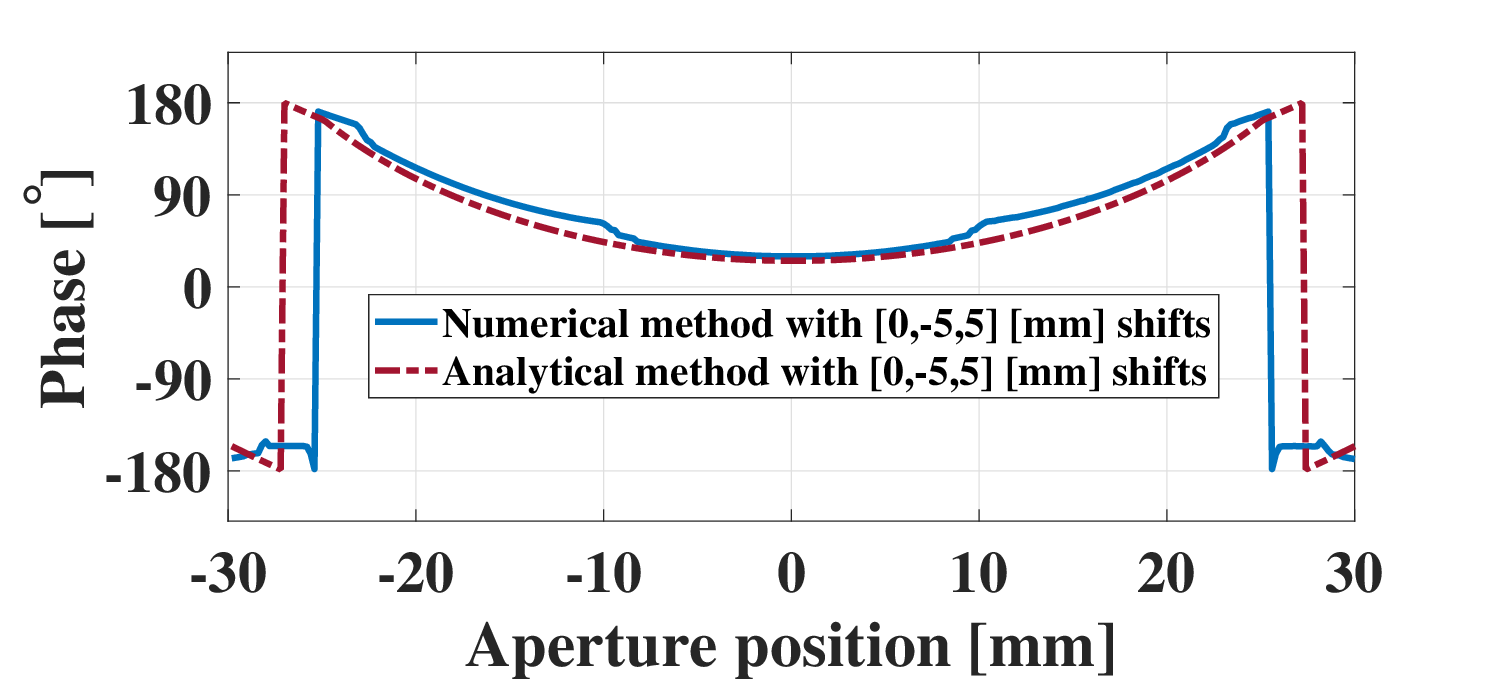}}
            
            \subfigure[]{\label{Fig: Fig04.c}
            \includegraphics[width=1\linewidth,trim={0.45cm 0.0cm 1.4cm 0.5cm},clip=true]{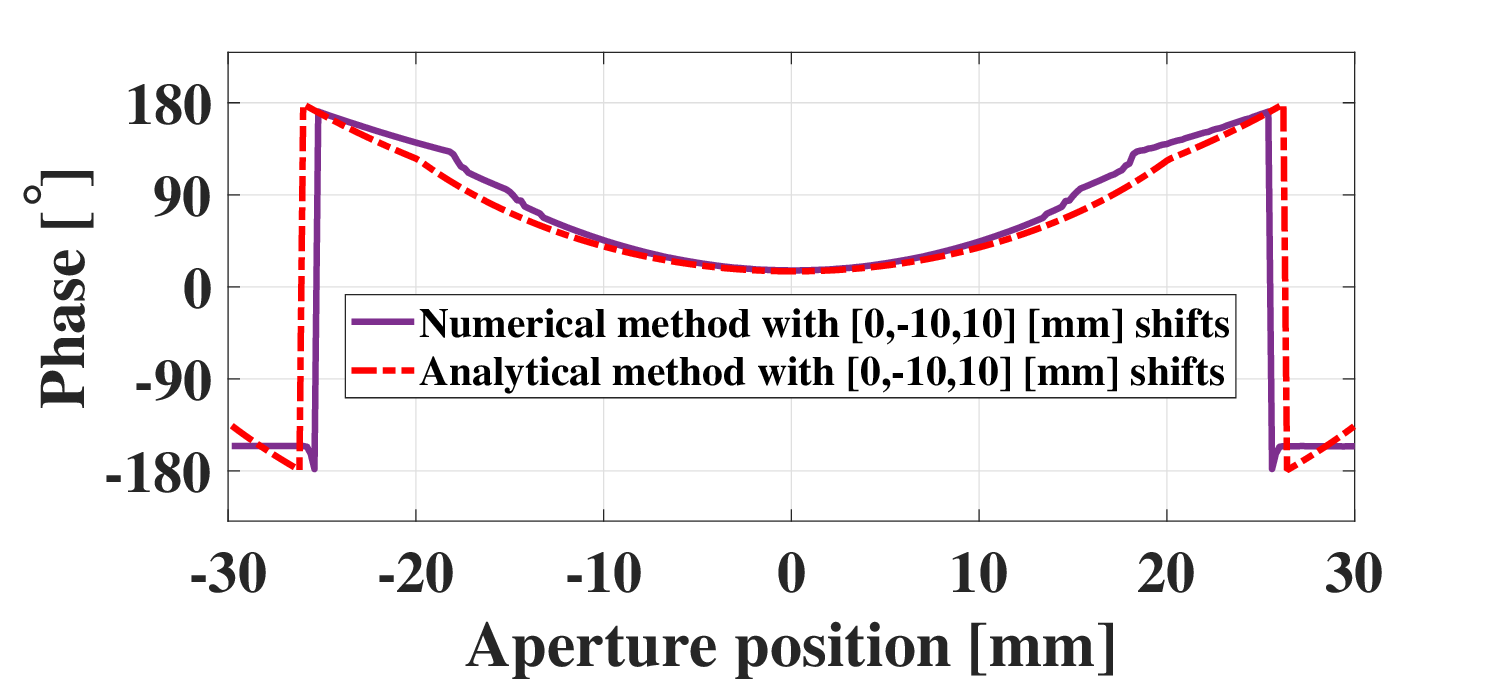}}
        \end{center}
        \vspace{-3.2mm}
        \caption{The analytical and numerical wrapped phase distribution generated by supercells for a 3-layer GRIN lens: (a) [0,0,0] [mm] shifts for the layers, (b) [0,-5,5] [mm] shifts for the layers, and (c) [0,-10,10] [mm] shifts for the layers.}
        \label{Fig: Fig04}
    \end{figure}

    Figure~\ref{Fig: Fig04} presents the phase delays generated by macrocells in the GRIN lens aperture, calculated using both analytical and numerical methods by transmission line theory and post-processing of the S-parameters, respectively. The figure demonstrates a rather good agreement between the analytical and numerical results, with a particular emphasis on the $[0,-5,+5]$~mm shift case.
    
\begin{figure}[ht!]
        \vskip-3mm
        \begin{center}
            \subfigure[]{\label{Fig: Fig05.a}
            \includegraphics[width=1\linewidth,trim={0,0cm 0,0cm 0,0cm 0,0cm},clip=true]{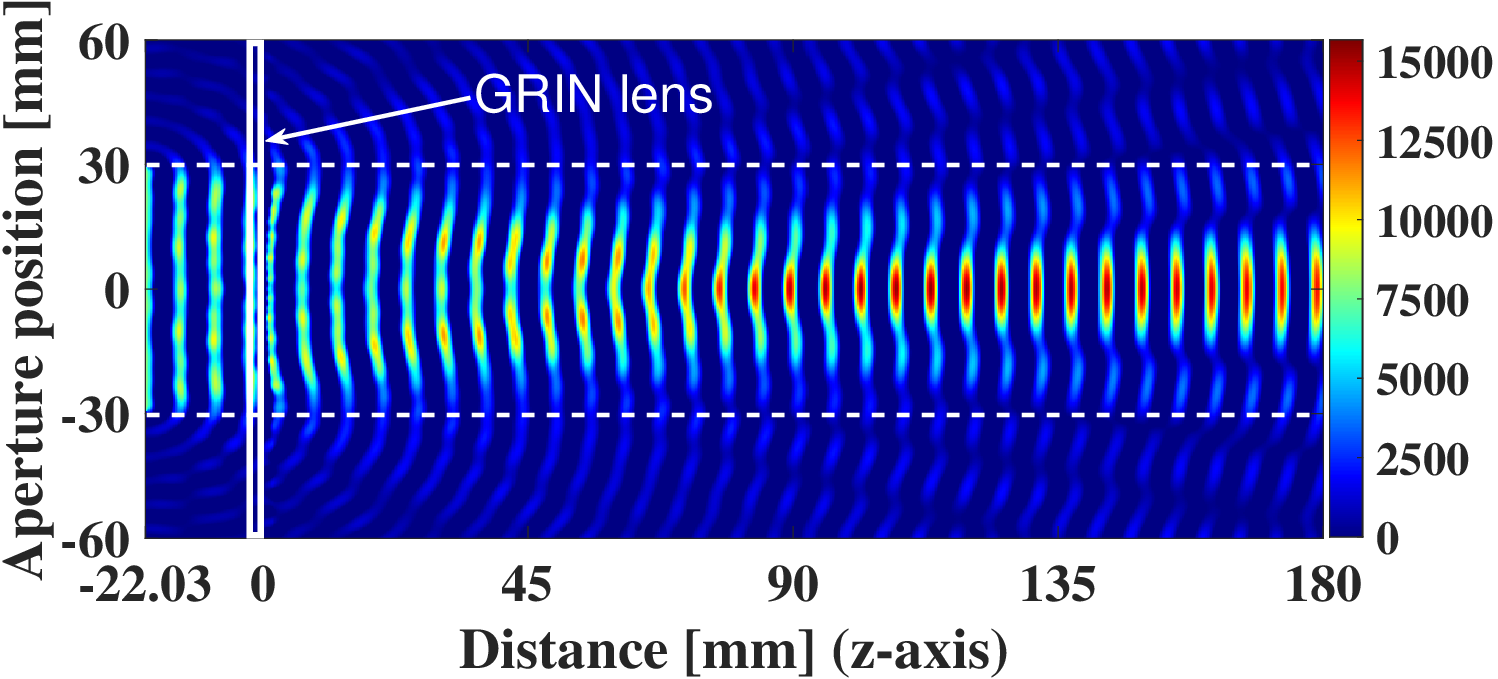}}
            
            \subfigure[]{\label{Fig: Fig05.b}
            \includegraphics[width=1\linewidth,trim={0,0cm 0,0cm 0,0cm 0,0cm},clip=true]{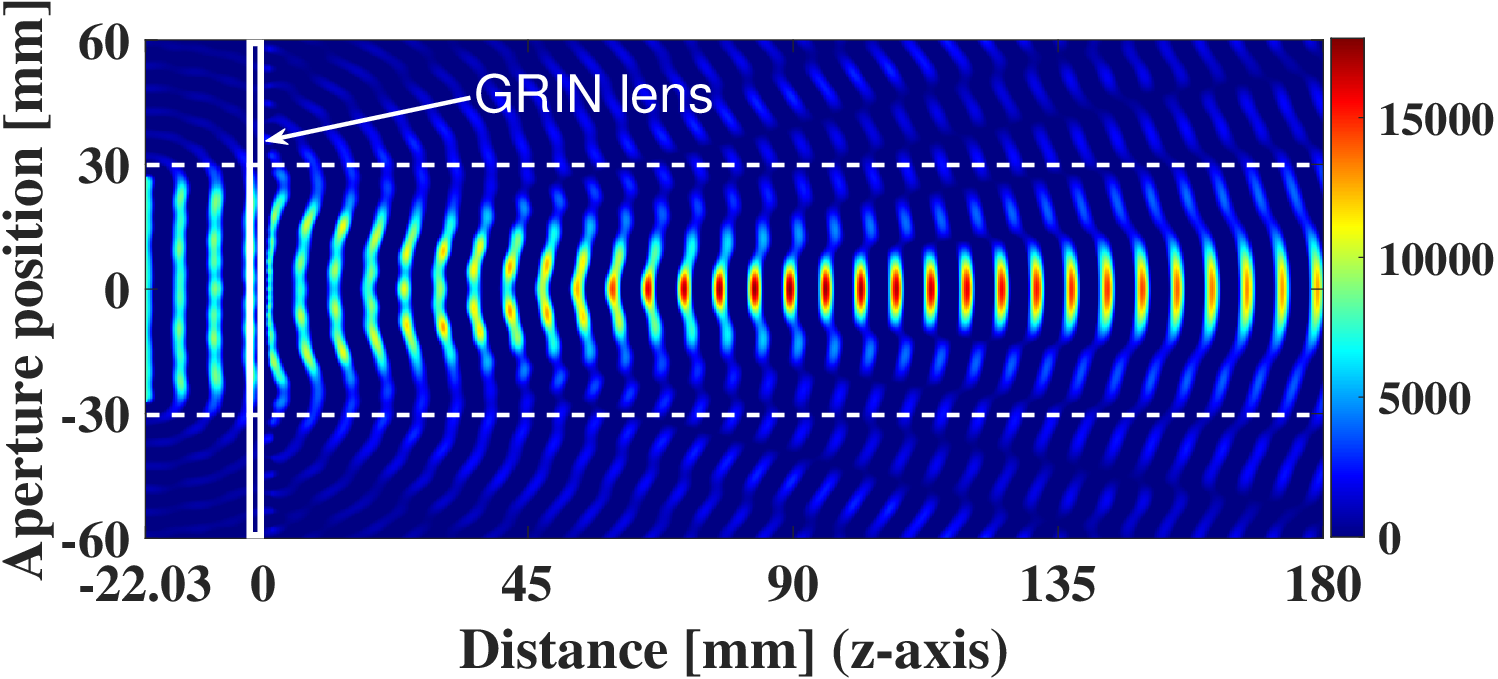}}
            
            \subfigure[]{\label{Fig: Fig05.c}
            \includegraphics[width=1\linewidth,trim={0,0cm 0,0cm 0,0cm 0,0cm},clip=true]{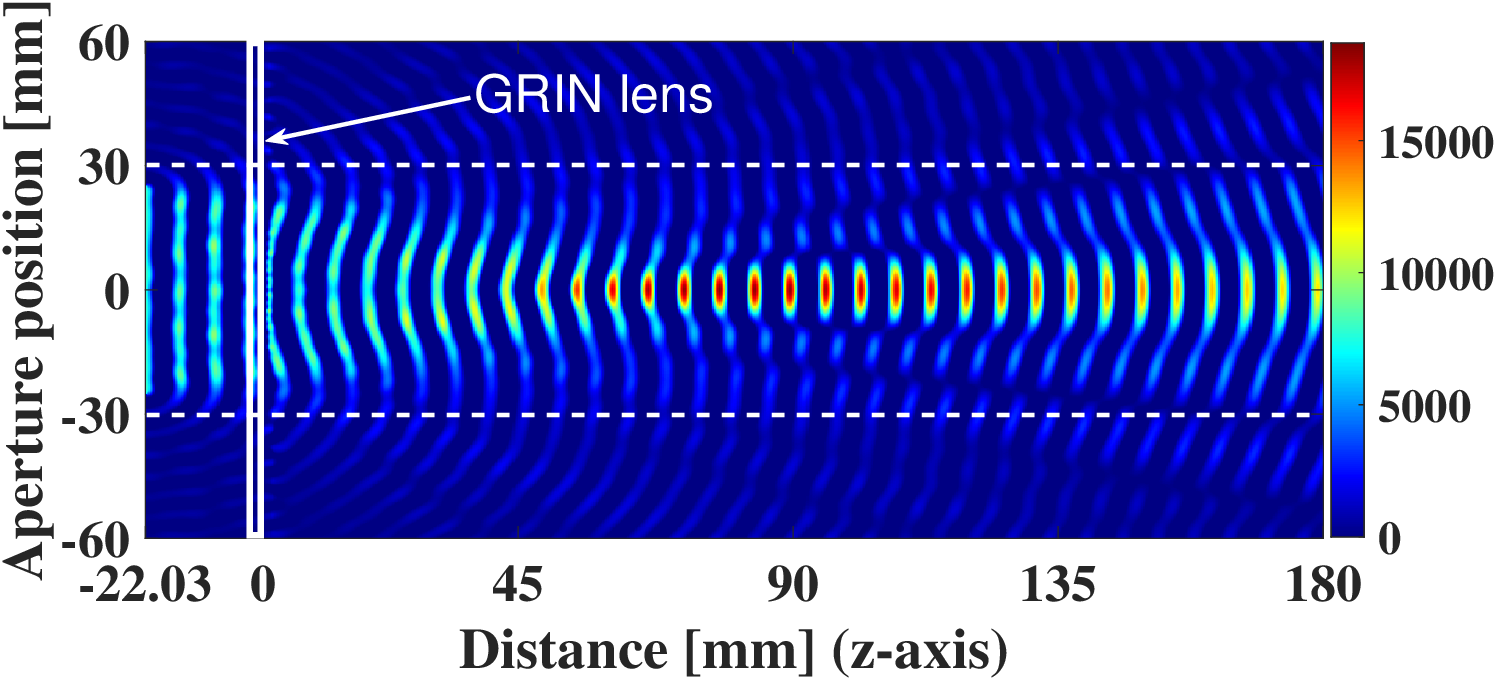}}
        \end{center}
        \vspace{-3.2mm}
        \caption{Electric field magnitude distribution in vicinity of focal points of a 3-layer GRIN lens, for varying layer shifts: (a) No shift; (b) $[0,-5,+5]$~mm shifts; (c) $[0,-10,10]$~mm shifts for the layers.}
        \label{Fig: Fig05}
    \end{figure}

    In Figs.~\ref{Fig: Fig05}(a-c), we depict the full-wave numerical results for the magnitude of the transmitted electric field, in which one can identify the points with maximum field intensity along the $z$-axis (i.e. the focal points) at about 110, 87, and 79~mm, respectively. It is evident from the figure that these focal points are offset from their expected positions. This discrepancy can be attributed to inaccuracies of the geometrical optics model when applied to a non-uniform structure such as the considered layered and patterned GRIN lens. Nevertheless, the results clearly demonstrate that the proposed reconfiguration method is operational.
    
    It is also worth noting that the cases related to $[0,-5,+5]$ and $[0,-10,+10]$~mm shifts exhibit two irregularities in their effective refractive index curves due to the shifts in the second and third layers and the truncation of basis function values, which is applied when creating the grooves in the layers.  Additionally, one may notice that the refractive index in the case of $[0,0,0]$~mm shift deviates from 1 at the aperture boundaries, Fig.~\ref{Fig: Fig03.b}. These irregularities lead to discontinuities in the gradient of the effective refractive index, which may also result in shifts of the focal points as compared to the expected positions.
    
    %
    
\section{Scalability Towards Infrared Applications}
\label{Scalability Towards Infrared Applications}
    While reconfigurable metalenses often struggle to balance high static performance with post-fabrication tunability, as the tuning mechanism typically deteriorates the original static performance, our GRIN-lens concept addresses this challenge by employing the second-order of first-kind Chebyshev polynomials for the static profile and the higher orders for reconfigurability. This approach minimizes perturbations, enabling tunable metalenses with high performance.
    
    Though currently demonstrated in microwaves, these concepts are readily extendable to shorter wavelengths, particularly the infrared region. The thermal-infrared spectrum, encompassing mid- and far-infrared regimes, supports crucial applications in sensing and imaging, including thermography across wide temperature ranges, chemical spectroscopy for molecular vibrational fingerprints, and remote sensing through atmospheric transmission windows~\cite{Baranov_2019, Li_2021}. However, while fields in Maxwell's equations are scalable, sources of mechanical vibrations are not, necessitating different material choices. From microwave to infrared frequencies, typical dielectrics exhibit lower refractive indices due to frequency dispersion, and metals experience increased ohmic losses, deviating from perfect electrical conductor behavior. Nevertheless, promising candidates exist, including $\mathrm{Ge}$, $\mathrm{SiO_2}$, $\mathrm{CaF_2}$, $\mathrm{ZnSe}$, and notably, chalcogenide phase-change compounds, which maintain high refractive indices and low losses extending into the far infrared~\cite{Kim_2024}. Manufacturing can also transition to nanofabrication methods, including optical lithography and nanoprinting, while dynamic modulation can be achieved through MEMS~\cite{Liu_2017}, tunable Joule heating~\cite{Liu_2022, Liu_2024}, and phase-change switching~\cite{Kim_2024}, suggesting promising prospects for expanding these lens methodologies to broader range of applications.
\section{Conclusion}
\label{Conclusion}
    In this work, we have introduced a reconfigurable GRIN lens for focusing and beamforming applications. The lens is layered along the optical axis, allowing for focal distance control by mechanical displacement of the layers (formed by different materials) along the lateral direction. Grooves are etched across the layers following profiles dictated by an appropriate functional basis to achieve the desired effective refractive index profiles. Selection of the basis functions, materials, and shifts for the layers is guided by an optimization algorithm. After an extensive evaluation, the first-kind Chebyshev polynomials have been chosen as basis functions. The analytical and numerical results have been obtained for a 3-layer reconfigurable GRIN lens with impedance-matched magneto-dielectric layers with the effective refraction indices similar to the ones for Rogers RO3006 and RO3010 substrates. The obtained analytical and numerical results clearly demonstrate the physical validity of the proposed reconfigurable GRIN lens concept. Our methodology is also readily scalable to the infrared regime, with proper adjustments in material selection and fabrication methods, enabling a much wider range of applications.
\section*{Acknowledgments}
    This work has been funded by FCT - Fundação para a Ciência e Tecnologia, I.P., by project with reference UIDB/50008/2020 and DOI identifier \href{https://doi.org/10.54499/UIDB/50008/2020}{10.54499/UIDB/50008/2020}. K.K. acknowledges co-financing by Fundação para a Ciência e a Tecnologia (Portuguese Foundation for Science and Technology) through the Carnegie Mellon Portugal Program under the fellowship PRT/BD/154201/2022. A.A. acknowledges financial support by Fundação para a Ciência e a Tecnologia (FCT) under PhD grant ref. 2022.13933.BD.
%
%
%
%
%

\end{document}